\documentclass[a4paper,12pt]{article}
\usepackage{amsfonts,slashed}
\usepackage{url}
\usepackage{latexsym}
\usepackage{amsfonts}
\usepackage{epsfig}
\usepackage{slashed}
\usepackage{latexsym,amssymb}
  \usepackage{amsmath,amssymb,amsthm}
\usepackage{ifpdf}
\ifx\pdfoutput\undefined
   \pdffalse
   \usepackage{cite}
 \else
   \pdfoutput=1
   \pdftrue
  \usepackage[pdftex]{hyperref}
\fi

\numberwithin{equation}{section}
\def\be{\begin{equation}}
\def\ee{\end{equation}}
\def\ba{\begin{array}}
\def\ea{\end{array}}

\newcommand{\bea}{\begin{eqnarray}}
\newcommand{\eea}{\end{eqnarray}}

\def\ii{{\rm i}}

\def\dM{{\partial\mathcal M}}

\newcommand{\bbox}{\lower.2ex\hbox{$\Box$}}

\def\bfone{\relax{\rm 1\kern-.35em 1}}
\def\bfzero{\relax{\rm I\kern-.18em 0}}

\begin{document}
\numberwithin{equation}{section}
\begin{flushright}
\today
\end{flushright}
\vskip 1cm
\begin{center}
{\bf\LARGE{Unconventional Supersymmetry at the Boundary of AdS$_4$ Supergravity}} \\
\vskip 2 cm
{\bf \large L.~Andrianopoli$^{1,2}$, B.~L.~Cerchiai$^{1,3}$,  R.~D'Auria$^{1}$,
 M.~Trigiante$^{1,2}$}
\vskip 8mm
 \end{center}
\noindent {\small $^1$ DISAT, Politecnico di Torino, Corso Duca
    degli Abruzzi 24, I-10129 Turin, Italy\\
    $^{2}$ Istituto Nazionale di
    Fisica Nucleare (INFN) Sezione di Torino, Italy\\
    $^3$ Centro Fermi -- Museo Storico della Fisica e Centro Studi e Ricerche ÒEnrico FermiÕ, Rome, Italy}

\vskip 2 cm

{\small \begin{center}{\bf Abstract}
\end{center}
In this paper we  perform,  in the spirit of the holographic correspondence, a particular asymptotic limit of  ${\mathcal N}=2$, AdS$_4$ supergravity  to ${\mathcal N}=2$ supergravity on a locally AdS$_3$ boundary. Our boundary theory enjoys ${\rm OSp}(2|2)\times {\rm SO}(1,2)$ invariance and is  shown to contain the $D=3$ super-Chern Simons ${\rm OSp}(2|2)$ theory  considered in \cite{Alvarez:2011gd} and featuring ``unconventional local supersymmetry". The model constructed in that reference describes the dynamics of  a spin-1/2 Dirac field in the absence of spin 3/2 gravitini and was shown to be relevant for the description of graphene, near the Dirac points, for specific spatial geometries.
Our construction yields the model in \cite{Alvarez:2011gd} with a specific prescription on the parameters. In this framework the  Dirac spin-1/2 fermion originates from the radial components of the gravitini in $D=4$.}
\vskip 1 cm

\vfill
\noindent {\small{\it
    E-mail:  \\
{\tt laura.andrianopoli@polito.it}; \\
{\tt bianca.cerchiai@polito.it}; \\
{\tt riccardo.dauria@polito.it}; \\
{\tt mario.trigiante@polito.it}}}
   \eject

\section{Introduction}
The AdS/CFT duality \cite{Maldacena:1997re} has provided important insights into the properties of strongly coupled non-gravitational models from a classical supergravity theory in one dimension higher. At low energies, in the supergravity limit of string theory,  it implies a one-to-one correspondence between quantum operators $\mathcal O$ in the boundary conformal field theory and fields $\phi$ of the bulk supergravity  and requires to supplement the supergravity action functional with appropriate boundary conditions $\phi^{(0)}$ for the supergravity fields, which act as sources for the operators of the CFT.
  As far as the metric field is concerned, in particular,  the bulk  metric is divergent near the boundary.  These divergences can be however disposed of successfully by the so called holographic renormalization \cite{holren} through the inclusion of appropriate counterterms at the boundary.

  In the spirit of the gauge-gravity correspondence the aim of this paper is to relate pure ${\mathcal N}=2$ AdS$_4$ supergravity to a $D=3$ Super-Chern-Simons theory developed in  \cite{Alvarez:2011gd}. This latter model displays ${\mathcal N}=2$ local supersymmetry in spite of the absence of gravitini. This special feature was named ``supersymmetry of a different kind" and later ``unconventional supersymmetry" by the authors. The only propagating field in the model is a spin-1/2 Dirac spinor with a possible mass term given in terms of the three-dimensional negative cosmological constant. The model discussed in \cite{Alvarez:2011gd} (denoted AVZ in the following) and further developed in \cite{Guevara:2016rbl, Alvarez:2013tga} was shown to have  important applications for some topics in condensed matter physics, in particular the description of graphene near the Dirac points, in agreement with the observation  that the special properties of this material can be formally investigated using the techniques of 3D gravity and quantum field theory \cite{Cortijo:2006ej, Iorio:2011yz, Cvetic:2012vg}.

In this paper we succeed in reproducing the AVZ $D=3$ model from ${\mathcal N}=2$, $D=4$ pure supergravity in the presence of a  3D boundary, using the results of  \cite{Andrianopoli:2014aqa} as a starting point. The  supersymmetry  of the $D=4$ supergravity lagrangian constrains the boundary values of the super field strengths to have definite values (Neumann boundary conditions), as was already noticed in \cite{Amsel:2009rr} in the ${\mathcal N}=1$ case.

The precise correspondence with the Chern-Simons model of AVZ  is found for a very specific choice of the $D=3$ boundary, corresponding to a local AdS$_3$ geometry placed at spatial infinity of the $D=4$ theory. Asymptotically AdS$_4$ solutions featuring this boundary geometry comprise the ``ultraspinning limit'' of  AdS$_4$-Kerr black hole considered in \cite{Caldarelli:2008pz, Caldarelli:2012cm, Gnecchi:2013mja, Klemm:2014rda, Hennigar:2015cja}.
 Once the curvatures and fields on the boundary are rewritten in terms of the 3D ones, in the appropriate limit they reproduce the field equations of \cite{Alvarez:2011gd},  provided, following the prescription of that paper, the $D=3$  gravitino only has a spin 1/2 part:
 \begin{equation}
 \psi_{A\,\mu }= \ii \gamma_\mu \chi_A\label{za}
 \,,
 \end{equation}
 where $A=1,2$, $\mu=0,1,2$ and $\gamma^\mu$ are $\gamma$-matrices in $D=3$.
{From} our point of view,  this assumption is  naturally embedded in the $D=4$ theory since it is compatible with the on shell condition on the $D=4$ gravitino $\Psi_{A\,\hat{\mu}}$ (here, $\hat\mu=0,1,2,3=(\mu,3)$, $A=1,2$ and we denote by $\Gamma^{\hat\mu}$ the $\gamma$-matrices in $D=4$):
\begin{equation}
\Gamma^ {\hat\mu}\Psi_{A\,\hat\mu }=0 \quad \Rightarrow\quad\Gamma^{\mu}\Psi_{A\,\mu} +\Gamma^{3}\Psi_{A\,3 }=0
\end{equation}
 which projects out its  spin-1/2 component in $D=4$ superspace, but still allows a non trivial solution
for $ \Gamma^{\mu}\Psi_{A\,\mu } $, if  the contribution from the radial component $\Psi_{A\,3} $ of the gravitino is not suppressed in the limit.


  The paper is organized as follows:

   In Sect. \ref{bADS4} we start by recalling, in subsections \ref{prelim} and \ref{bou}, the main facts about the general formulation of  pure ${\mathcal N}=2$ AdS$_4$ supergravity as developed in \cite{Andrianopoli:2014aqa}, where the boundary terms compatible with a full supersymmetry invariance of the supergravity action are computed in a geometric approach.
  In particular in subsections \ref{ADS3description} and \ref{asymptoticlimit} the conditions on the field strengths of the various $D=4$ fields at the boundary are written in a covariant form with respect to Lorentz group at the boundary. Referring to the Fefferman-Graham parametrization of an asymptotically AdS four-dimensional space-time ${\mathcal M}$, a suitable behavior of the fields near the boundary $\dM$   is devised in order to obtain the model in \cite{Alvarez:2011gd}. To this aim we use a description of the asymptotically AdS$_4$ geometry which is typical of the ``ultraspinning" limit of the AdS-Kerr solution. It is shown that, in the boundary limit $r\rightarrow \infty$, the conditions of the field strengths, which are required by $D=4$ supersymmetry, naturally yield the field equations of a $D=3$ super gravity Lagrangian with symmetry ${\rm OSp}(2|2)\times {\rm SO}(1,2)$ which, as it is
 known, can be  formulated in terms of (super) Chern-Simons theories~\cite{Achucarro:1987vz, Achucarro:1989gm, Witten:1988hc, Witten:2007kt, Witten:2003ya, Cacciatori:2005wz, Leigh:2008tt, Townsend:2013ela}.

  In Sect.~\ref{SODK}, which is the central part of our paper, we explicitly compare the Chern-Simons lagrangian of \cite{Alvarez:2011gd} with the $D=3$ supergravity lagrangian whose field equations do coincide with those obtained from the asymptotic values on $\dM$ of the super field-strengths.  We show that, by imposing the condition (\ref{za}) on the limit values on $\dM$ of the fields and on the $D=3$ Lagrangian, we exactly reproduce the theory of ref.~\cite{Alvarez:2011gd} as a part of the $\mathcal{N}=2$ super-AdS$_3$ supergravity  Lagrangian.

  Sect.~\ref{newinsights} contains a discussion of our results and the specific features of the $D=3$ models that we obtain from its holographic relation with $D=4$ supergravity. We deduce the explicit relations among the  fields naturally associated with the boundary limit of the $D=4$ super vielbein and spin connection, and the corresponding fields of ref.~\cite{Alvarez:2011gd}, where instead supersymmetry is an internal symmetry leaving invariant the space-time dreibein. In particular we find that the supersymmetry parameter of the $D=3$ theory is proportional to the propagating spinor field of the AVZ model. Furthermore,  from the correspondence with $D=4$ supergravity, we gain an interpretation of
 the propagating spinor field of the AVZ model  in terms of the radial component of the $D=4$  gravitino.

  Our results and possible further developments are presented in Sect.~\ref{Conclusion}.

  We left to the Appendices  some technical material, including the  map between our notations and the ones of \cite{Alvarez:2011gd}, the details of the ultraspinning limit and the explicit relation between the space-time dreibein of \cite{Alvarez:2011gd} and the superdreiben on $D=3$ superspace that we get in the limit.
\section{Boundary behavior of ${\mathcal N}=2$, AdS$_4$ supergravity  } \label{bADS4}
\subsection{Preliminaries}\label{prelim}
In the presence of a space-time defect, such as a boundary of space-time, compatibility of the lagrangian with local supersymmetry invariance requires appropriate  boundary conditions to be imposed. This was first pointed out in \cite{York:1972sj}, \cite{Gibbons:1976ue}, in  early attempts to study the quantization of gravity with a path integral approach, in order to have an action which depends only on the first derivatives of the metric.  More recently, the addition of boundary terms was considered in \cite{Horava:1996ma} to cancel gauge and gravitational anomalies in the Ho\v{r}ava--Witten model in 11D.
Inclusion of boundary terms and counterterms is also an essential tool for the study of the AdS/CFT duality \cite{Maldacena:1997re} and, as far as the bosonic sector of AdS supergravity is concerned, has been extensively studied in many different contexts.   In particular, interesting results were obtained in  \cite{Aros:1999id}, \cite{Miskovic:2009bm}, where it was shown that the addition of the topological Euler--Gauss-Bonnet term to the Einstein action of four dimensional AdS gravity leads to a background-independent definition of Noether charges, without imposing  Dirichlet boundary conditions on the fields.
  Such boundary term indeed regularizes the action and the related (background independent) conserved charges.

At the full supergravity level, boundary contributions  were considered from several authors, using different approaches, and in  particular in \cite{Amsel:2009rr}, \cite{vanNieuwenhuizen:2005kg}, \cite{esposito}, \cite{Moss:2003bk}, \cite{Howe:2011tm}, and more recently in \cite{Freedman:2016yue}, \cite{Genolini:2016ecx}, \cite{Drukker:2017xrb}.
 In \cite{vanNieuwenhuizen:2005kg} the minimal $AdS$ supergravity theory in three dimensions with a boundary and with no boundary conditions on the fields was constructed from the requirement of local supersymmetry invariance of the supergravity action,  \emph{without imposing Dirichlet boundary conditions on the fields}, in contrast to the Gibbons--Hawking prescription \cite{Gibbons:1976ue}.  Within that  approach,  it was shown in particular that ${\mathcal N}=1$, $D=3$ pure supergravity, including its appropriate boundary term, actually reproduces not only the Gibbons-Hawking-York boundary term, but also the counterterm which regularizes the total action, in the language of holographic renormalization. The problem of boundary conditions in the holographic framework was discussed in generality in \cite{Amsel:2009rr}, in particular for the minimal, ${\mathcal N}=1$, AdS$_4$ supergravity.

\subsection{Supersymmetric boundary conditions}
\label{bou}

In \cite{Andrianopoli:2014aqa}, the results of \cite{Aros:1999id, Aros:1999kt}, \cite{Mora:2004kb}, \cite{Olea:2005gb}, \cite{Jatkar:2014npa} on four dimensional gravity were extended to the supersymmetric case for ${\mathcal N}=1$ and ${\mathcal N}=2$  pure supergravity, in the geometric approach, by adding to the supergravity lagrangian in four dimensions, besides the  Euler--Gauss-Bonnet contribution, other topological terms allowing full supersymmetry invariance of the supergravity action, in the bulk and on the boundary of four dimensional space-time. Supersymmetry invariance constrains the boundary values of the superfield-strengths (Newmann boundary conditions), without fixing however  the superfields themselves on the boundary.
In particular, in \cite{Andrianopoli:2014aqa} a pure $D=4,\,\mathcal{N}=2$ theory was written in the presence of boundary. If we denote by $\mathcal{M}$ the four-dimensional space-time and by $\partial\mathcal{M}$ its three-dimensional boundary, supersymmetry requires the following equations to be satisfied at the boundary $\partial\mathcal{M}$ (still written in terms of four dimensional fields):\footnote{
 We shall use the notations of \cite{Andrianopoli:2014aqa}, where in particular the metric is mostly minus.
 \\
 With respect to that paper, however, here we made  some changes which make the formulas more transparent and better adapted to match the notations in three dimensions. More precisely, the spin connection and the curvature are defined with an extra minus sign: $\omega^{ab}\rightarrow -\omega^{ab},\,R^{ab}\rightarrow -R^{ab}$, and the 1-form  with a prefactor $-\frac{1}{\sqrt{2}}$ :$A \rightarrow -\frac{1}{\sqrt{2}}\,A$. We will use Majorana spinors both in four as well in three dimensions and redefine the constants appearing in the quoted paper as follows:
\begin{equation}\label{relat}
\frac 1\ell=2e=\frac{P}{\sqrt 2}=\sqrt{ -\frac{\Lambda}{3}}\,;\quad L=\frac{1}{\sqrt 2}\,,
\end{equation}
where $\Lambda$  is the cosmological constant and  $\ell$  is the AdS$_4$ radius.}
\begin{align}
{R}^{ab}|_{\partial\mathcal M}&=\left[\frac 1{\ell^2}\,V^a\wedge V^b+\frac 1{2\ell}\,\bar{\Psi}_A\Gamma^{ab}\wedge\Psi_A\right]_{\partial\mathcal M}\,,\label{AdSbound0}\\
 \nabla^{(4)}V^a|_{\partial\mathcal M}&=\frac{i}{2}\,\bar{\Psi}_A\Gamma^{a}\wedge\Psi_A|_{\partial\mathcal M}\,,\label{AdSbound1}\\
dA^{(4)}|_{\partial\mathcal M}&=\bar{\Psi}_A\wedge\Psi_B\,\epsilon_{AB}|_{\partial\mathcal M}\,,\label{AdSbound2}\\
 \nabla^{(4)}\Psi_A|_{\partial\mathcal M}&=  \frac {\ii}{2\ell}\, \Gamma_a\Psi_A\,\wedge V^a|_{\partial\mathcal M}\,,\label{AdSbound3}
\end{align}
where $a,b,...=0,1,2,3$ are four dimensional flat indices, $ \nabla^{(4)}$ denotes the $\rm{SO}(1,3)$ covariant differential, and we have defined:
\begin{align}
{R}^{ab}&\equiv d\omega^{ab}+\omega^a{}_c\wedge \omega^{cb}\,,\nonumber\\
 \nabla^{(4)}V^a&\equiv dV^a +\omega^a{}_b\wedge V^b\,,\nonumber\\
 \nabla^{(4)}\Psi_A&\equiv  d\Psi_A+\frac{1}{4}\,\omega_{ab}\,\Gamma^{ab}\wedge\Psi_A-\frac 1{2\ell}\,\epsilon_{AB}\,A^{(4)}\,\wedge\Psi_B\,.
\end{align}
In what follows, for the sake of notational simplicity, we shall omit the wedge symbol.\par
Equations (\ref{AdSbound0})-(\ref{AdSbound3}) express the condition of an asymptotic super-AdS$_4$ geometry.

\subsection{Explicit $D=3$ description}\label{ADS3description}

The explicit $D=3$ description of the above supergravity equations depends on the general symmetry properties of the theory on the boundary $\dM$ which we wish to relate to.

Our aim in the present paper is to derive the model in \cite{Alvarez:2011gd}, which features locally AdS$_3$ geometries, as the effective  theory on an asymptotic boundary $\dM$, placed at $r\to \infty$. To this end we need to suitably choose the boundary behavior of the $D=4$ fields, which relate them to the $D=3$ ones. \par
We refer to a Fefferman-Graham parametrization of the four-dimensional geometry, where the coordinates $x^{\hat\mu}$ of the asymptotically-AdS$_4$ geometry split into the three coordinates $x^\mu$, $\mu=0,1,2$ on $\dM$ and the radial coordinate $x^3=r$. The boundary is located at $r\rightarrow \infty$.\par
 Let us start rewriting the curvatures in (\ref{AdSbound0})-(\ref{AdSbound3}) in a ${\rm SO}(1,1)\times {\rm SO}(1,2)$-covariant way, where ${\rm SO}(1,1)$ acts as a dilation on the radial variable $r$ and ${\rm SO}(1,2)$ is identified with the Lorentz group on $\dM$. Given the extrinsic curvature 1-form $ h^i =\omega^{3i}$, we introduce new 1-forms $K_+^i,\,K_-^i$, as follows \footnote{Since we use $a,b,\dots=0,1,2,3$ as rigid $D=4$ Lorentz indices, we denote the three dimensional ${\rm SO}(1,2)$ ones by $i,j,k,\dots=0,1,2$.}:
\begin{align}
K_+^i\equiv \frac{1}{2}\left(V^i-\ell\,\omega^{3i}\right)\,\,,\,\,\,K_-^i\equiv \frac{1}{2}\left(V^i+\ell\,\omega^{3i}\right)\,.\label{defef}
\end{align}
These 1-forms are 3-vectors with respect to ${\rm SO}(1,2)$-Lorentz group on the boundary with different ${\rm SO}(1,1)$-grading: $K_+^i$ have grading $+1$ and $K_-^i$ grading $-1$. Similarly we decompose the four-component  gravitinos into two-component ones corresponding to different  ${\rm SO}(1,1)$-gradings:
\begin{equation}
\Psi_{A}=\Psi_{+\,A}+\Psi_{-\,A}\,\,,\,\,\,\,\,\Gamma^3\Psi_{\pm\,A}=\pm \ii \,\Psi_{\pm\,A}\,.
\end{equation}
In terms of these new quantities we find that equations (\ref{AdSbound0})-(\ref{AdSbound3}) on the boundary $\dM$ can be cast in the  following form:
\begin{align}
\mathcal{R}^{ij}&\equiv d\omega^{ij}+\omega^i{}_k\wedge \omega^{kj}=\frac{4}{\ell^2}\,K_+^{[i}\wedge K_-^{j]}+\frac{1}{\ell}\,\bar{\Psi}_{+\,A}\Gamma^{ij}\Psi_{-\,A}\,,\nonumber\\
dV^3&=-\omega^3{}_i\wedge V^i+\bar{\Psi}_{+\,A}\Psi_{-\,A}\,,\nonumber\\
{\mathcal{D}}K_\pm^i&\equiv  dK_\pm^i+\omega^{i}{}_j\wedge K_\pm^j=\frac{\ii}{2}\,\bar{\Psi}_{\pm A}\Gamma^i\Psi_{\pm A}\pm\frac{1}{\ell}\,K_\pm^i\wedge V^3\,,\nonumber\\
{\mathcal{D}}{\Psi}_{\pm A}&\equiv d{\Psi}_{\pm A}+\frac{1}{4}\omega_{ij}\,\Gamma^{ij}\wedge{\Psi}_{\pm A}=
- \frac{\ii}{\ell}\,K_{\pm i}\,\wedge\Gamma^i{\Psi}_{\mp A}+\frac{1}{2\ell}\,\epsilon_{AB}\,A^{(4)}\wedge {\Psi}_{\pm |B}\pm \frac{1}{2\ell}\,{\Psi}_{\pm A}\wedge V^3\,,\nonumber\\
dA^{(4)}&=\epsilon_{AB}\bar{\Psi}_{A}\,\wedge\Psi_{B}=2\,\epsilon_{AB}\bar{\Psi}_{+A}\,\wedge
\Psi_{-B}\,\label{curvs3}
\end{align}
where use has been made of equations (\ref{app1}) of Appendix~\ref{Notations} and we have identified
 the rigid four-dimensional index $a=3$ with the radial direction orthogonal to $\dM$ at the boundary: $V^3=dV^3=0$. This in particular implies  the conditions \footnote{One can easily show that the extrinsic curvature 1-form $\omega^3{}_i$ cannot have a component along the gravitino field of the superspace.}:
\begin{equation}\label{dv3}
\omega^3{}_i = h_{i|j}\,V^j\,\mbox{  with } h_{i|j}=h_{j|i}\,; \quad \bar{\Psi}_{+\,A}\Psi_{-\,A}=0\,.
\end{equation}

\subsection{Asymptotic limit}\label{asymptoticlimit}

Since the model in \cite{Alvarez:2011gd} which we want to relate to is defined on locally AdS$_3$ geometries, we shall use a description of the asymtptic behavior for the $D=4$ metric and the spin-connection corresponding to an asymptotically locally AdS$_4$ space-time with a locally AdS$_3$ boundary. An example of such  backgrounds is the ``ultraspinning limit'' of an AdS$_4$-Kerr black-hole defined in  \cite{Caldarelli:2008pz, Caldarelli:2012cm}, see discussion in Appendix~\ref{Aultraspinning}. \footnote{See also the solution found in \cite {Klemm:2014rda} which is described, in a suitable limit, by the same metric.} In particular we choose the following boundary behavior for the $D=4$ fields \footnote{The ensuing  boundary behavior is related to the that given in \cite{Amsel:2009rr} by the change of coordinates given in Appendix~\ref{Aultraspinning}.}:
\begin{align}\label{beha}
K_+^i(x,r)&=\frac{r}{\ell}\,E^i(x)+\dots\,\,,\,\,\,K_-^i(x,r)=\frac{1}{4}\,\frac{\ell}{r}\,E^i(x)+\dots\,\,,\,\,\,V^3(r)=\frac{\ell}{r}\,dr+\dots\,,\nonumber\\
\Psi_{+\,A\, \mu}(x,r)&=\sqrt{\frac{r}{\ell}}\,(\psi_{A\,\mu},{\bf 0})+\dots\,\,,\,\,\,\,\Psi_{-\,A\,\mu}(x,r)=\sqrt{\frac{\ell}{r}}\,\left({\bf 0},\,\frac{\varepsilon}{2}\, \psi_{A\,\mu}\right)+\dots\,,\nonumber\\
\omega^{ij}(x,r)&=\omega^{ij}(x)+\dots\,\,\,,\,\,\,\,\,\,
A^{(4)}(x,r)=\varepsilon \, A_\mu(x^\nu)\,dx^\mu+\dots\,.
\end{align}
where $\psi_{A\,\mu}=(\psi_{A\,\mu\,\alpha})$, $\mu=0,1,2$, $\alpha=1,2$, are gravitini in $D=3$, associated with Majorana spinor 1-forms $\psi_A\equiv\psi_{A\,\mu} dx^\mu$ and $\varepsilon=\pm 1$. The ellipses refer to subleading terms in the  $r\rightarrow\infty$ limit. The boundary behavior of the radial components of the $D=4$ gravitini, $\Psi_{\pm A \,3}$, will be explicitly discussed later, in  eq. (\ref{psir}) of Section~\ref{newinsights}.
  In order to make contact with the model of \cite{Alvarez:2011gd}, we shall assume that the $O\left(\frac{\ell^3}{r^3}\right)$  term in the expansion of $V^i$ is absent. This further restriction is satisfied by an AdS$_3$ slicing of AdS$_4$ space-time, see Appendix~\ref{Aultraspinning}, obtained by sending, in the ultraspinning AdS-Kerr solution, the mass of the black hole to zero.\footnote{In fact it is satisfied by a more general space-time which is the BTZ \cite{Banados:1992wn, Banados:1992gq} slicing of a locally AdS$_4$, known as `BTZ- black string' and found in \cite{Emparan:1999fd}, see Sect.~\ref{Conclusion}.} It implies the vanishing of the energy momentum tensor of the boundary theory  \cite{Amsel:2009rr,Caldarelli:2012cm}(Neumann condition):
  $$T_{ij}=0\,,$$
   which is a feature of the model of \cite{Alvarez:2011gd},\cite{Guevara:2016rbl} that we want to make contact with.
   By supersymmetry this also implies restrictions on the subleading terms in the expansion of the gravitinos \cite{Amsel:2009rr,Caldarelli:2012cm}.
In order to reproduce the effect of deformations of this model one should consider more general boundary behaviors. We leave this to a future investigation.\par
The components $\omega^{ij}$, restricted to $\dM$, provide the spin-connection on that space, and $\mathcal{R}^{ij}$ the corresponding Riemann curvature 2-form. Let us emphasize that the fields $E^i,\,\psi_A,\,\omega^{ij},\,A$ are to be intended, in the spirit of the geometric approach to supergravity, as superfields in the superspace extension of $\dM$, corresponding to the (non-dynamical) super-dreibein and gauge field on $\dM$.\par

 If we now perform the $r\to \infty$ limit, the leading contributions to (\ref{curvs3}) correspond to the following equations on the boundary:
\begin{align}
\mathcal{R}^{ij}&=\frac{1}{\ell^2}\,E^{i}  E^{j}+\frac{\varepsilon}{2\ell}\,\bar{\psi}_A\gamma^{ij}\psi_A\,, \nonumber\\
{\mathcal{D}}E^i&=\frac{\ii}{2}\,\bar{\psi}_{A}\gamma^i\psi_{A}\,, \label{asymptoticeq} \\
{\mathcal{D}}\psi_A&=-\varepsilon\,
\frac{\ii}{2\ell}\,E_i\,\gamma^i  \psi_A+\frac{\varepsilon}{2\ell}\,\epsilon_{AB}\,A \,  {\psi}_{B}\,,\nonumber\\
dA&=\epsilon_{AB}\bar{\psi}_{A}\psi_{B}\,.\nonumber
\end{align}
where $\mathcal D$ denotes Lorentz covariant derivative in $D=3$, as defined in (\ref{curvs3}). It is expressed in terms of the (super) torsionless  spin connection $\omega^{ij}$. \footnote{As it is usual in supergravity and evident from the second equation in (\ref{asymptoticeq}), the spin connection $\omega^{ij}$ has instead a non-zero space-time torsion, which is proportional to a gravitino bilinear. } They define the  vanishing of the supercurvatures of  ${\mathcal N}=2$ super AdS$_3$ algebra.

The set  of constraints (\ref{asymptoticeq}) on the 3D boundary can be found  as equations of motion from the following $D=3$ supergravity lagrangian  \cite{Achucarro:1987vz}:
\begin{equation}\label{lag3D}
 \mathcal{L}^{(3)}= \left(\mathcal R^{ij} -\frac{1}{3\,\ell^2}  E^i E^j  -\frac{\varepsilon}{2\,\ell} \, \bar\psi_A\gamma^{ij}\psi_A\right) E^k \epsilon_{ijk} - \frac{\varepsilon}{2\,\ell} AdA +2 \bar\psi_A\left(\mathcal{D} \psi_A - \frac{\varepsilon}{2\ell}\,\epsilon_{AB}\,A \,  {\psi}_{B}\right)\,.
\end{equation}
Eq. (\ref{lag3D}) collects in fact two inequivalent lagrangians on $D=3$ superspace, depending on the sign $\varepsilon=\pm 1$.  As found in \cite{Achucarro:1987vz},
 in  $D=3$ a family of inequivalent ${\mathcal N}$-extended supergravity lagrangians can be constructed, enjoying invariance under ${\rm OSp}(p|2) \times {\rm OSp}(q|2) $, with $p+q={\mathcal N}$. In our case, where ${\mathcal N}=2$, the family of inequivalent lagrangians consists of only two elements, that we labeled with the sign $\varepsilon$. They are invariant under ${\rm OSp}(2|2)_{(\varepsilon)} \times {\rm SO}(1,2)_{(-\varepsilon)}$. For the time being we keep  both theories, which differ by a flipping of the gravitino gauge charge. In the next section we will show that one of the two assignements (in particular, $\varepsilon=-1$) reproduces the results of \cite{Alvarez:2011gd}.

In the geometric formalism, the supersymmetry transformation of a generic field $\Phi$, with supersymmetry parameter $\epsilon_A$,  can be found as a super-diffeomorphism in the fermionic directions of superspace:
\begin{equation}\label{K}
 \delta \Phi= \imath_K(d\Phi)+ d(\imath_K \Phi) \, ,
\end{equation} generated by the tangent vector $K=\bar\epsilon_A D^A$, where $D^A$ is the tangent vector dual to the gravitino field, such that $\psi_A(D^B)=\delta_A^B$.
The lagrangian (\ref{lag3D}) is then invariant under the following supersymmetry transformations:
\begin{align}\label{susytr}
\delta \omega^{ij}&=\frac{\varepsilon}{\ell}\,\bar{\epsilon}_A\gamma^{ij}\psi_A\,, \nonumber\\
\delta E^i&=\ii \,\bar{\epsilon}_{A}\gamma^i\psi_{A}\,,\\
\delta\psi_A&={\mathcal{D}}\epsilon_A\,+\,\varepsilon\,
\frac{\ii}{2\ell}\,E_i\,\gamma^i  \epsilon_A-\frac{\varepsilon}{2\ell}\,\epsilon_{AB}\,A \,  {\epsilon}_{B}\,,\nonumber\\
\delta A&=2\epsilon_{AB}\bar{\epsilon}_{A}\psi_{B}\,. \nonumber
\end{align}

The full $\mathfrak{osp}(2|2)_{(\varepsilon)} \times {\mathfrak{so}}(1,2)_{(-\varepsilon)}$ can be made manifest by introducing the torsionful spin connections
\cite{Achucarro:1989gm}:
\begin{equation}\label{omegavareps}
\omega_{(\pm\varepsilon)}^{ij} =  \omega^{ij} \pm \frac{\varepsilon}{\ell} \, E_k \epsilon^{ijk}=\epsilon^{ijk}\,\omega_{\pm(\varepsilon)k}\,,
\end{equation}
 in terms of which the constraints (\ref{asymptoticeq}) read:
\begin{align}
\mathcal{R}^{i}_{(\varepsilon)}&=\ii\, \frac{\varepsilon}{\ell}\,\bar{\psi}_A\gamma^{i}\psi_A\,, \label{asymptoticRip}\\
\mathcal{R}^{i}_{(-\varepsilon)}&=0 \,,\label{asymptoticRim}\\
{\mathcal{D}_{(\varepsilon )}}\psi_A&=\frac{\varepsilon}{2\ell}\,\epsilon_{AB}\,A   {\psi}_{B}\,,\label{asymptoticpsi}\\
dA &= \epsilon_{AB}\bar{\psi}_{A}\psi_{B}\, , \label{asymptoticA}
\end{align}
where we defined:
\begin{eqnarray}
 R^i_{(\pm)} &\equiv& d\omega^i_{(\pm)} -\frac 12 \omega^j_{(\pm)}\omega^k_{(\pm)}\varepsilon^i{}_{jk}
\end{eqnarray}
In terms of the new connections, the supersymmetry transformations (\ref{susytr}) become:
\begin{align}\label{susytrman}
 \delta \omega^i_{(-\varepsilon)}=0\,, \quad \delta \omega^i_{(\varepsilon)}=\varepsilon\,\frac{2\ii}{\ell}\,\bar{\epsilon}_A\gamma^{i}\psi_A\,,\quad
\delta A = 2\epsilon_{AB}\bar{\epsilon}_{A}\psi_{B}\,, \nonumber\\
\quad\delta\psi_A={\mathcal{D}_{(\varepsilon)}}\epsilon_A\,-\frac{\varepsilon}{2\ell}\,\epsilon_{AB}\,A \,  {\epsilon}_{B}\equiv \nabla^{(\varepsilon)}\epsilon_A\,.
\end{align}
As it is evident from eq.s (\ref{susytrman}), in this theory the superalgebra $\mathfrak{osp}(2|2)\times\mathfrak{so}(1,2)$ (and in particular supersymmetry) is realized as a gauge symmetry, since eq.s (\ref{asymptoticRip}) - (\ref{asymptoticA}) can be interpreted as the vanishing of the supercurvatures of the above superalgebra (isomorphic to ${\mathcal N}=2$ super AdS$_3$).

As shown in \cite{Achucarro:1989gm}, using (\ref{omegavareps}) the supergravity lagrangian (\ref{lag3D}) can be written as
\begin{equation}
\label{SUGRA3D}
\mathcal{L}^{(3)} =\varepsilon\left( \mathcal{L}_{(\varepsilon)} - \mathcal{L}_{(- \varepsilon)}\right)\,\equiv \mathcal{L}_+^{(3)}-\mathcal{L}_-^{(3)}
\end{equation}
 where:
 \begin{align}
 \mathcal{L}_{(\varepsilon)}&=\,
\frac \ell 2\left( \omega_{(\varepsilon)}^i d\omega_{(\varepsilon)|i} -\frac 13  \omega_{(\varepsilon)}^i\omega_{(\varepsilon)}^j\omega_{(\varepsilon)}^k\varepsilon_{ijk} \right)
+2\varepsilon \,\bar \psi_A \nabla^{{(\varepsilon)}} \psi_A -\frac{\varepsilon}{2\ell} A\,dA \,,\label{supercs}\\
 \nabla^{(\varepsilon)} \psi_A&\equiv\,(d + \frac 14 \omega_{(\varepsilon)}^{ij} \gamma_{ij})\,\psi_A -\frac{\varepsilon}{2\ell} A \,\psi_B \,\epsilon_{AB} \,,\\
 \mathcal{L}_{(-\varepsilon)}&=\,
\frac \ell 2\left( \omega_{(-\varepsilon)}^i d \omega_{(-\varepsilon)|i} -\frac 13  \omega_{(-\varepsilon)}^i\omega_{(-\varepsilon)}^j\omega_{(-\varepsilon)}^k\varepsilon_{ijk} \right) \,.
 \end{align}
 The component
 $\mathcal{L}_{(\varepsilon)}$ is the super-Chern Simons lagrangian of the superalgebra $\mathfrak{osp}(2|2)_{(\varepsilon)}$, while  $\mathcal{L}_{(-\varepsilon)}$ is the  Chern Simons lagrangian of the  algebra $\mathfrak{so}(1,2)_{(-\varepsilon)}$.

 \section{Comparison with ``Supersymmetry of a different kind"}\label{SODK}
 In  ref.~\cite{Alvarez:2011gd}, the $D=3$ Chern-Simons action for a ${\rm OSp}(2|2)$ connection was considered, and shown to be invariant off-shell under gauge supersymmetry.
 We aim here to compare it with the boundary theory discussed above. The explicit map between our notations and  conventions on the space-time signature and fields  with the ones of \cite{Alvarez:2011gd} is given in Appendix~\ref{mapnot}.

A peculiar feature of the ${\rm OSp}(2|2)$ 1-form connection in \cite{Alvarez:2011gd} is that the spinor 1-form associated with the odd generator of the superalgebra is not a spin-3/2 gravitino, but is  instead given in terms of a (${\mathcal N}=2$) spin $1/2$ field that we name $\chi_A$,\footnote{In ref.~\cite{Alvarez:2011gd} the spin-1/2 Dirac field is called $\psi$. Since in the supergravity literature the symbol $\psi$ is generally reserved to the gravitino, for the sake of clarity we preferred to adhere to the latter convention. Note that the extra factor i with respect to the conventions of ref. \cite{Alvarez:2011gd} ensures that $\chi_A$ is Majorana, being related to the Majorana gravitino 1-form $\psi_A$. Referring to the notation in \cite{Alvarez:2011gd}, the spinors $\ii \chi_A $  introduced here should therefore be identified with the real and imaginary part of the spin-$\frac 12$ fermion $\psi$ of \cite{Alvarez:2011gd}:  $\psi=\chi_1+ \ii \chi_2$.} defined by the following  condition
\begin{equation}\label{gravchi}
  \psi_A=\ii \,e_i\,\gamma^i\chi_A\,,
\end{equation}
where $e_i$ is a 1-form space-time dreibein, \emph{invariant} under supersymmetry transformations:
\begin{equation}
 \delta e^i = 0\,,\label{deltae0}
\end{equation}
and satisfying:
\begin{equation}
\hat{\mathcal{D}}e^i=\frac{1}{\ell}\,\epsilon^{ijk}\,e_j\, e_k\,,\label{dZ}
\end{equation}
where $\hat{\mathcal{D}}$ denotes a covariant derivative with respect to an appropriate torsionful connection $\omega^i$.

Once expressed in terms of the same notations, the super Chern-Simons lagrangian introduced in Section 3 of \cite{Alvarez:2011gd} does coincide, modulo an overall scaling, with our lagrangian (\ref{supercs}) \emph{for the choice $\varepsilon=-1$}, if one identifies the spin-connection $\omega^i$ of \cite{Alvarez:2011gd} with our $\omega_{(\varepsilon)}^i=\omega_{(-)}^i$.
For later convenience, in the following we shall write the lagrangian and the ensuing equations of motion in terms of a generic choice of  $\varepsilon=\pm 1$.
Indeed, if we plug the Ansatz (\ref{gravchi}) in the lagrangian (\ref{supercs}) we obtain:
\begin{align}
 \mathcal{L}_{(\varepsilon)}\,=&
\frac \ell 2\left( \omega_{(\varepsilon)}^i d \omega_{(\varepsilon)|i} -\frac 13  \omega_{(\varepsilon)}^i\omega_{(\varepsilon)}^j\omega_{(\varepsilon)}^k\varepsilon_{ijk} \right)+\,2\varepsilon \,e_i\,\mathcal{D}^{(\varepsilon)}e^i\,\bar{\chi}_A\chi_A+ \nonumber\\&- 4\, \ii \,\,\varepsilon \,\bar \chi_A \slashed{\nabla}^{{(\varepsilon)}} \chi_A\, {\rm e}\,d^3x\,
-\frac{\varepsilon}{2\ell} A\,dA \,,\label{supercs2}
\end{align}
where ${\rm e}\equiv {\rm det}(e_\mu{}^i)$ and we have defined:
 \begin{equation}
 \slashed{\nabla}^{{(\varepsilon)}} \chi_A\equiv \gamma^i\,{\nabla}^{{(\varepsilon)}}_i \chi_A=
 \slashed{\mathcal{D}}^{(\varepsilon)}\chi_A -\frac \varepsilon{2\ell}\,A_i \epsilon_{AB}\gamma^i\chi_B\,.
 \end{equation}
 In deriving the above lagrangian we have used the following properties:
\begin{equation}
\bar{\psi}_A{\nabla}^{(\varepsilon)}\psi_A=e_i\,\mathcal{D}^{(\varepsilon)}e^i\,\bar{\chi}_A\chi_A- 2\, \ii\, {\rm e}\,\bar \chi_A \slashed{\nabla}^{{(\varepsilon)}} \chi_A\,d^3x\,,
\end{equation}
being $e^i\wedge e^j\wedge e^k=\epsilon^{ijk}\,{\rm e}\,d^3x$. Note that, defining the Dirac spinor $\chi=\chi_1+\ii\,\chi_2$, we can also rewrite the kinetic term for the two spinors $\chi_A$ as:
\begin{equation}
\bar{\chi}_A \slashed{\mathcal{D}}^{{(\varepsilon)}} \chi_A=\frac{1}{2}\,\bar\chi \left(\overrightarrow{\slashed{\partial}}-\overleftarrow{\slashed{\partial}}-
\frac{1}{2}\,\gamma^i\slashed{\omega}_{ij}\,\gamma^j
\right)\chi\,,
\end{equation}
which is useful for comparison with \cite{Alvarez:2011gd}.
The  equation of motion for the spinors $\chi_A$ is the Dirac equation
\footnote{In terms of the Dirac spinor $\chi$ we can write the above equation in the same form as \cite{Alvarez:2011gd}:
\begin{equation}
\left(\ii\slashed{\partial}+\frac{k}{2}-\frac{\ii}{4}\gamma^i\slashed{\omega}_{ij}\,\gamma^j-\frac{\varepsilon}{2\ell}
\slashed{A}\right)\chi+\frac{\ii}{2{\rm e}}\partial_\mu\left(e_i{}^\mu\,\gamma^i\right)\,\chi=0\,,
\end{equation}
where $e_i{}^\mu e_\mu{}^j=\delta_i^j$.}
\begin{eqnarray}
\slashed{\nabla}^{(\varepsilon)}\chi_A  -\frac{\ii}2\,\kappa \chi_A  &=& 0  \,,\label{diracchi}
\end{eqnarray}
where  $\kappa$ is defined as $e_i\mathcal{D}^{(\varepsilon)}e^i=-\kappa\,{\rm e}\,d^3x$, as in \cite{Alvarez:2011gd}, and in our case $\kappa=\varepsilon\,6/\ell$.
Eq. (\ref{diracchi}) matches with the equation obtained from (\ref{asymptoticpsi}) by means of (\ref{gravchi}).
Indeed,   from (\ref{asymptoticpsi}), using (\ref{gravchi}) and (\ref{dZ}), we easily derive the following equation:
\begin{equation}
\nabla^{(\varepsilon)}_i\chi_A=\frac{\ii \varepsilon}{\ell}\,\gamma_i\chi_A\,,\label{Dchigchi}
\end{equation}
which, when contracted with $\gamma^i$, yields the Dirac equation (\ref{diracchi}). As shown in  \cite{Guevara:2016rbl},  equation (\ref{Dchigchi}) ensures that no local spin-$3/2$ components are generated through parallel transport of the fermions.

We remark that, as explicitly shown in Appendix \ref{proof}, eq.
 (\ref{Dchigchi}) implies
\begin{equation}\label{dxx}
  d(\bar\chi_A\chi_A)=0\,,
\end{equation}
so that $\bar\chi\chi$ is a free constant of the model. This is a consequence of having fixed the Weyl invariance hidden in (\ref{gravchi}) in such a way that, in agreement with ref. \cite{Alvarez:2011gd}, the torsion tensor in (\ref{dZ}) only has its fully antisymmetric component.
\par
The supersymmetry transformation of the spinor $\chi_A$ can be found from (\ref{susytr}) and (\ref{deltae0}), given the relation (\ref{gravchi}), and reads:
\begin{equation}\label{deltachi}
\delta \chi_A = -\frac{\ii}{3}
\gamma^i\,\nabla_i^{(\varepsilon)} \epsilon_A\,.
\end{equation}
Consistency requires $\nabla_i^{(\varepsilon)} \epsilon_A$ to have no spin-$3/2$ component, see \cite{Alvarez:2011gd}:
\begin{equation}
P_i{}^j\,\nabla^{(\varepsilon)}_j\epsilon_A=0\,,\label{projepsilon}
\end{equation}
where $P_i{}^j=\delta_i^j-\frac{1}{3}\,\gamma_i\gamma^j$ is the projector on the spin-$3/2$ component, which is a condition on $\epsilon_A$. In next section we shall find that our derivation of the model implies a relation between $\epsilon_A$ and $\chi_A$ which is consistent with (\ref{projepsilon}) by virtue of Eq. (\ref{Dchigchi}), in accordance with the fact that supersymmetry is a global symmetry of the $D=3$ model, which does not affect the counting of the degrees of freedom.
\footnote{We thank Pedro Alvarez and Jorge Zanelli for a relevant comment on this point.}
\par

The equations of motion for $\omega^{i}_{(\varepsilon)}$ and $A$ following from the lagrangian (\ref{supercs2}) are given by
\begin{align}\label{3Deq}
\mathcal{R}^{i}_{(\varepsilon)}&=- \frac{\varepsilon}{\ell}\,\bar{\chi}_A \chi_A \epsilon^{ijk} e_j e_k\,, \\
dA &= i \epsilon_{AB} \bar{\chi}_{A} \gamma^k \chi_{B}  e^i e^j \epsilon_{ijk}\,.
\end{align}
They reproduce exactly those obtained respectively from the equations (\ref{asymptoticRip}), (\ref{asymptoticA}), substituting the Ansatz (\ref{gravchi}).
\par
The equation of motion of the $e^i$ field implies the vanishing of the energy momentum tensor $T_{ij}=0$ \cite{Guevara:2016rbl}.  \footnote{Care must be taken to obtain this result from (\ref{SUGRA3D}) since it contains a torsionful connection, so that  the contribution from the contorsion tensor  to the matter Lagrangian should be included.}
This is consistent with the assumption we made in Section \ref{asymptoticlimit} when defining the asymptotic limit of the $D=4$ fields.
Indeed, as shown in  \cite{Amsel:2009rr,Caldarelli:2012cm}, the absence of $\mathcal{O}\left(\frac{\ell^3}{ r^3}\right)$ terms in the expansion of the vielbein is related to the Neumann condition $T_{ij}=0$.

\section{New insights from the $D=4$ approach}\label{newinsights}
After having derived the model in \cite{Alvarez:2011gd} as a holographic limit of AdS$_4$ supergravity, let us now emphasize some specific features of our construction of the three-dimensional theory within the four-dimensional supergravity.
\begin{itemize}
\item{Our boundary lagrangian is $\mathcal{L}^{(3)}$ in (\ref{lag3D}) and is built from the difference of $\mathcal{L}_{(\varepsilon)}$ and $\mathcal{L}_{(-\varepsilon)}$, as in (\ref{SUGRA3D}). This expresses the fact that our $D=3$ boundary model always has an ${\rm OSp(2|2)_{(\varepsilon)}\times SO(1,2)_{(-\varepsilon)}}$ symmetry. In particular the cosmological constant of the three dimensional space-time, which defines the mass $|\kappa|$ of the Dirac spinor $\chi$, as observed in  \cite{Alvarez:2011gd}, in our case is always negative (local AdS$_3$ geometry).}
\item{The space-time dreibein $e^i$ introduced in \cite{Alvarez:2011gd} for the definition  (\ref{gravchi}) satisfies:
\begin{equation}
\mathcal{D}_{(\varepsilon)} e^i\equiv de^i-\epsilon^{ijk}\omega_{(\varepsilon)\,j}\, e_k=-\frac{\varepsilon}{\ell}\,\epsilon^{ijk}\,e_j\, e_k\,.\label{d+e}
\end{equation}
 It should  not to be confused with the superfield dreibein $E^i$, appearing in (\ref{asymptoticeq}).
  The presence of two sets of  dreibein, $e^i$ and $E^i$, satisfying different differential equations, seems at first sight a puzzling feature of our model.
To clarify this point, let us explicitly compare the torsion equation for the  (non dynamical) bosonic dreibein $e^i$, eq. (\ref{d+e}), with the super-torsion equation in (\ref{asymptoticeq}) for the super-dreibein $E^i$ entering the definition of $\omega^i_{(\pm)}$,  after using (\ref{gravchi}).
If  written  in terms of the spin connection $\omega^i_{(\varepsilon)}$, the second and third equation in (\ref{asymptoticeq}) read:
\begin{eqnarray}
\mathcal{D}_{(\varepsilon)} E^i\equiv dE^i-\epsilon^{ijk}\omega_{(\varepsilon)\,j}\, E_k&=&-\frac{\varepsilon}{\ell}\,\epsilon^{ijk}\,E_j\, E_k + \frac 12 \bar\chi_A \chi_A \epsilon^{ijk} e_j \,e_k\,,\label{d+E}\\
\gamma_i(\mathcal{D}_{(\varepsilon)}\chi_A)e^i+ \gamma_i\chi_A\mathcal{D}_{(\varepsilon)}e^i&=&\frac \varepsilon{2\ell}A \epsilon_{AB}\gamma_i\chi_B e^i\,.\label{d+chie}
\end{eqnarray}
Eq. (\ref{d+E}) contains in its right hand side a redundant set of vielbein, and hints to a relation expressing the $E^i$ in terms of the basis of space-time dreibein $e^i$ and of the complex spinor  $\chi\equiv \chi_1 +\ii \chi_2$.

We consider the following Ansatz:
\begin{eqnarray}\label{ansE}
E^i &=& \left(1 + a\,  \bar{\chi}\chi + b (\bar{\chi}\chi)^2\right) \,e^i\,,
\end{eqnarray}
Plugging (\ref{ansE}) in (\ref{d+E}), and using (\ref{d+e}), (\ref{d+chie}), one can determine $E^i$ to be:
\begin{eqnarray}\label{ansEdet}
E^i &=& \left(1 +\varepsilon \,\frac\ell 2\,  \bar{\chi}\chi -\frac{\ell^2}4 (\bar{\chi}\chi)^2\right) \,e^i\equiv M(\bar\chi\chi) \,e^i\,.
\end{eqnarray}
The details of the calculation are given in Appendix~\ref{proof}.

Let us now check compatibility of the supersymmetry transformation of $E^i$, in eq. (\ref{susytr}), with the one obtained by its expression (\ref{ansEdet}), using (\ref{deltachi}). The direct comparison yields:
\begin{eqnarray}\label{condE}
  \bar\epsilon_A(\gamma^{ij}+\eta^{ij})\chi_A &=& \frac 23 \ii\, \eta^{ij} (a+2b \bar{\chi}\chi)\bar{\chi}_A \gamma^k\nabla^{(\varepsilon)}_k \epsilon_A\,,
\end{eqnarray}
that in particular requires $\bar\epsilon_A\gamma^{ij}\chi_A =0$. This condition can be non trivially satisfied if:
\begin{eqnarray}
  \epsilon_A &=& \ii N(\bar\chi\chi)\,\chi_A\,, \quad N(\bar\chi\chi)= \alpha(x)+\beta(x) \bar\chi\chi
\end{eqnarray}
with $\alpha,\beta$ real functions of $D=3$ space-time.
Substituting this Ansatz in (\ref{condE}), we find the solution:
\begin{eqnarray}
\alpha=0\,,\quad  \epsilon_A &=& \ii \beta \,\bar\chi_B \chi_B \chi_A\,,\quad \beta \mbox{ constant}
\end{eqnarray}
which still depends on the free  parameter $\beta $.
}
\item{The reinterpretation of the model of \cite{Alvarez:2011gd} in the framework of AdS$_4$ supergravity gives a new perspective on the meaning of the spinor $\chi$.
 Indeed, the condition (\ref{gravchi}) identifies the spinor $\chi_A$ as the spin-$\frac 12$ projection of the $D=3$ gravitino $\psi_A$:
\begin{equation}\label{chi}
\chi_A= -\frac \ii 3 \,\gamma_i\, e^{i|\mu}\psi_{A\,\mu} \,,
\end{equation}
where $\mu=0,1,2$ labels curved space-time indices. In our framework, however, we are  expressing $\psi_A$ in terms of the ${\mathcal N}=2$ gravitino in $D=4$, $\Psi_{A\,\hat\mu}$, where $\hat{\mu}
=(\mu,3)$. The gravitino in $D=4$ obeys the condition
\begin{equation}\label{gra4}
  \Gamma_a V^{a|\hat{\mu}}\Psi_{A\,\hat\mu}=0\,,
\end{equation}
which projects out its spin-$\frac 12$ component all over $D=4$ superspace, including its boundary $\dM$.\par
Expanding Eq. (\ref{gra4}) in the limit $r\rightarrow \infty$ and using the boundary behavior (\ref{beha}) for the supergravity fields, eq. (\ref{gra4}) implies the following behavior for the radial component $\Psi_{A\,3}$ of the gravitino fields:
\begin{equation}\label{psir}
\psi_{+\,A\,3}=\frac{3\ii\varepsilon}{2M(\bar \chi \chi)}\,\left(\frac{\ell}{r}\right)^{\frac{5}{2}}\,(\chi_A,{\bf 0})\,\,,\,\,\,\,\,\psi_{-\,A\,3}=\frac{3\ii}{M(\bar \chi \chi)}\,\left(\frac{\ell}{r}\right)^{\frac{3}{2}}\,({\bf 0},\chi_A)\,.
\end{equation}
 In this way we give an interpretation of the spin $1/2$ fields $\chi_A$ as originating from the radial component of the gravitino field on $\dM$. These components, although subleading with respect to the others, are related to them in a non-trivial way through Eq. (\ref{gra4}).}
\end{itemize}

\section{Conclusions and outlook} \label{Conclusion}
In this paper we have found a holographic relation, at the full supersymmetric level, between AdS$_4$, $\mathcal{N}=2$ pure supergravity and a $D=3$ Chern-Simons theory, found in \cite{Alvarez:2011gd}, exhibiting unconventional supersymmetry, which was shown to describe the behavior of graphene near the Dirac points \cite{graphene}.
It is well known that there exists a close analogy  between the  properties of some condensed matter special systems, in particular of graphene, and the theoretical framework of quantum field theory and general relativity \cite{Cortijo:2006ej},\cite{Iorio:2011yz},\cite{Cvetic:2012vg}.
  The  formalisms of these theories can and have been  used in the past years because the equations of motion of the quasiparticles in graphene (at the Dirac points) have formally the same expression as the Dirac equation of the free electrons  both in special and general relativity. In particular, the powerful methods of the relativistic theories turn out to be a very useful  tool  for exploring the special properties of graphene possessing a two-dimensional spatially curved surface. Viceversa graphene  has been proposed  as a simple laboratory to check gravitational cosmic phenomena like Hawking-Unruh radiation etc.

 In our paper  the model of reference   \cite{Alvarez:2011gd} was reproduced, after projecting out the spin 3/2 part of the $D=3$ gravitino, as a theory living at the boundary of $\mathcal{N}=2$, $D=4$ Supergravity, in the spirit of AdS$_4$/CFT$_3$ correspondence (for a recent review on a top-down approach to the subject, see \cite{Semenoff:2012xu}).
This  hints to conjecture  a more physical meaning for such relationship. Supersymmetry plays a crucial role in establishing the above correspondence, and
the results presented in this paper reinforce the holographic
 relationship between (2+1)-dimensional condensed matter models and supergravity at the full supersymmetric level, already in the absence of interactions.

Consistently with \cite{Alvarez:2011gd}, the $D=3$ model that we derive has vanishing energy-momentum tensor, as it follows from the equations of motion for $e_\mu{}^i$. The analyses of \cite{Amsel:2009rr,Caldarelli:2012cm} imply that the dual supergravity background is locally AdS.  The duality correspondence discussed in  the present paper refers to ``vacuum'' theories.
Reproducing the effects of deformations of the CFT on $\dM$ calls for a generalization of boundary behaviors (\ref{beha}).
Already in \cite{Alvarez:2011gd} it was observed that the solutions of the model include the Ba\~nados-Teitelboim-Zanelli black holes \cite{Banados:1992wn, Banados:1992gq}. In light of our duality with $D=4$ supergravity, these solutions correspond to the so-called `BTZ- black string' found in \cite{Emparan:1999fd}:
\begin{equation}
ds^2=\left(1+\frac{\rho^2}{\ell^2}\right)\,ds^2_{BTZ}-\frac{d\rho^2}{1+\frac{\rho^2}{\ell^2}}\,,
\end{equation}
where $ds^2_{BTZ}$ is the metric of a generic BTZ black hole and we can identify $\rho=\ell\,\sinh(\alpha)$.
The above issues are left to  future work.

\par
Finally, let us observe that the Supergravity approach gives, as an extra bonus, the possibility to obtain directly a gauge supersymmetry in (2+1)-dimensional models, even if the Lagrangian of the mother theory only enjoys a supersymmetric invariance which is not gauge since its algebra only closes on-shell.

\section*{Acknowledgements}

The motivation of this paper arose from a stimulating and fruitful discussion with Jorge Zanelli, to whom we express all our gratitude.
\\
We also acknowledge interesting discussions with Pedro Alvarez, Andr\'es Anabalon, Matteo Baggioli, Antonio Gallerati, Dietmar Klemm and Lucrezia Ravera.

\appendix
\section{Notations and conventions} \label{Notations}
Throughout the paper we used a mostly minus signature, both in four and  in three dimensions.  World indices are denoted with greek letters: $\hat\mu...=0,1,2,3=(\mu,3)$; $\mu,...=0,1,2$; $x^3=r$ denoting the radial coordinate, while to label flat indices we used latin indices: $a,...=0,1,2,3=(i,3)$; $i,...=0,1,2$; the flat direction orthogonal to $\dM$ being labeled with 3.
\subsection{Conventions on spinors in $D=4$ and $D=3$}
It is convenient to use the following representation of the Clifford algebra:
\begin{align}
\Gamma^i&=\left(\begin{matrix}{\bf 0} & \gamma^i\cr \gamma^i & {\bf 0}\end{matrix}\right)=\sigma_1\otimes \gamma^i\,\,,\,\,\,\gamma^0=\sigma_2,\,\gamma^1=i\sigma_1,\,\gamma^2=i\sigma_3\,,\nonumber\\
\Gamma^3&=\left(
\begin{array}{cccc}
 i & 0 & 0 & 0 \\
 0 & i & 0 & 0 \\
 0 & 0 & -i & 0 \\
 0 & 0 & 0 & -i
\end{array}
\right)=i\,\sigma_3\otimes {\bf 1}\,,\nonumber\\
\Gamma^5&=i\,\Gamma^0\Gamma^1\Gamma^2\Gamma^3=\left(
\begin{array}{cccc}
 0 & 0 & i & 0 \\
 0 & 0 & 0 & i \\
 -i & 0 & 0 & 0 \\
 0 & -i & 0 & 0 \\
\end{array}
\right)=-\sigma_2\otimes {\bf 1}\,.\label{Laurabasis}
\end{align}
The charge conjugation matrix is taken to be
\begin{equation}
C=\Gamma^0\,\,,\,\,\,C^{-1}\Gamma^a\,C=-(\Gamma^a)^T\,.
\end{equation}
Let us define  $\Psi_\pm$ as the following projections:
\begin{equation}
\Gamma^3\Psi_\pm=\pm i \,\Psi_\pm\,.\label{proj1}
\end{equation}
A Majorana spinor $\lambda$ satisfies the condition:
\begin{equation}
\bar{\lambda}\equiv {\lambda}^\dagger \Gamma^0={\lambda}^T\,C \Longrightarrow \lambda=\lambda^*\,,
\end{equation}
namely it is real.
In the new basis the projected Majorana spinor 1-forms $\Psi_\pm$ satisfying the condition $\Gamma^3\Psi_\pm=\pm i \,\Psi_\pm$ have the following forms:
\begin{equation}
\Psi_+=(\xi_1,\xi_2,0,0)\,\,,\,\,\Psi_-=(0,0,\xi_3,\xi_4)\,\,\,\,,\,\,\,\,\xi_\alpha^*=\xi_\alpha\,.
\end{equation}
The following bilinears vanish:
\begin{equation}
\overline{\Psi_\pm}\Gamma^{ij}\Psi_\pm=\overline{\Psi_\pm}\Gamma^3\Psi_\pm=\overline{\Psi_\pm}\Psi_\pm=0\,.
\label{app1}
\end{equation}
\subsection{Map between our notations in $D=3$ and the ones in \cite{Alvarez:2011gd}} \label{mapnot}
The conventions used here are quite different from the ones of \cite{Alvarez:2011gd} in various respects. For the sake of clarity, we are providing here the relations between the notions used throughout this paper and the ones of \cite{Alvarez:2011gd}. To make the comparison, the items referring to \cite{Alvarez:2011gd} are labeled with a tilde.
\begin{enumerate}
  \item[i)]
  We are using a mostly minus signature for the space-time metric, $\eta = (+,-,-)$, while $\tilde{\eta}=(-,+,+)$ is mostly plus. they are related by
  $$\tilde\eta_{ij}=-\eta_{ij}\,,$$
from which it descends:
\begin{align}\label{tildegamma}
 \tilde\gamma^i & = \ii \gamma^i\,\,,\,\,\, \tilde\gamma_i  =- \ii \gamma_i\,,\nonumber\\
 \tilde{\epsilon}_{ijk}&=\epsilon_{ijk}\,\,,\,\,\,\tilde{\epsilon}^{ijk}=-\epsilon^{ijk}\,,\nonumber\\
 \tilde{\omega}^i{}_j&={\omega}^i{}_j\,\,,\,\,\, \tilde{\omega}_{ij}=-{\omega}_{ij}\,\,,\,\,\, \tilde{\omega}^{i}=\frac{1}{2}\,\tilde{\epsilon}^{ijk}\,\tilde{\omega}_{jk}=\frac{1}{2}\,{\epsilon}^{ijk}\,{\omega}_{jk}=\omega^i\,,\nonumber\\
 \tilde{J}_i&=-\frac{\tilde{\gamma}_i}{2}= \ii \,\frac{{\gamma}_i}{2}=J_i\,.
\end{align}
  \item[ii)] In  \cite{Alvarez:2011gd} all fields are part of a dimensionless connection. The dimensions of the $D=3$ supergravity fields considered here are:
  \begin{align}
  [\psi_A]=[L^{\frac{1}{2}}]\,\,,\,\,\, [E^i]=[e^i]=[L]\,\,,\,\,\, [\chi_A]=[L^{-\frac{1}{2}}]\,\,,\,\,\, [A]=[L]\,\,,\,\,\, [\omega^{ij}]=[L^0]\,.
  \end{align}
Let us denote here by $\tilde{\chi}$ the spin-$1/2$ field in \cite{Alvarez:2011gd}, denoted there by $\psi$. The relation between the fields used here and in \cite{Alvarez:2011gd}
is:
\begin{align}
\tilde{e}^i&={e}^i\,\,,\,\,\,\tilde{\chi}=-\frac{1}{\sqrt{\ell}}\,\chi=-\frac{1}{\sqrt{\ell}}\,(\chi_1+\ii\,\chi_2)\,\,,\,\,\,
\overline{\tilde{\chi}}=-\frac{1}{\sqrt{\ell}}\,\bar{\chi}\,\,,\,\,\,\,\tilde{A}=\frac{A}{2\ell}\,.
\end{align}

\end{enumerate}

\section{The ultraspinning Kerr-AdS black hole}\label{Aultraspinning}
In this appendix we give a specific four-dimensional background which satisfies the boundary conditions (\ref{beha}).
It is the ultraspinning limit of the four-dimensional Kerr-AdS geometry \cite{Caldarelli:2008pz, Caldarelli:2012cm, Gnecchi:2013mja, Klemm:2014rda, Hennigar:2015cja}. This space-time is obtained by
performing the limit $a\rightarrow \ell$ ($\ell$ being the AdS radius) of the rotation parameter $a$ of a Kerr-AdS black hole, keeping the horizon area fixed and, at the same time, zooming into the pole. The metric reads:
\begin{equation}
ds^2=V(\rho)\,\left[dt-\ell\,\sinh^2\left(\frac{\sigma}{2}\right)\,d\phi\right]^2-\frac{\ell^2+ \rho^2}{4 }\,\left(d\sigma^2+\sinh^2(\sigma) d\phi^2\right)-\frac{d\rho^2}{V(\rho)}\,,\label{ultraspinning}
\end{equation}
where:
\begin{equation}
V(\rho)=-\frac{2 M \rho}{ \rho^2+\ell^2}+\frac{\rho^2}{\ell^2}+1\,,
\end{equation}
The boundary is located at $\rho\rightarrow \infty$  and its metric $ds^2_{{\rm bdry}}$ is:
\begin{equation}
ds^2_{{\rm bdry}}=\lim_{\rho\rightarrow \infty} \frac{\ell^2}{\rho^2}\,ds^2=\left[dt-\ell\,\sinh^2\left(\frac{\sigma}{2}\right)\,d\phi\right]^2-
\frac{\ell^2}{4}\left(d\sigma^2+\sinh^2(\sigma)\,d\phi^2\right)\,.
\end{equation}
It describes an AdS$_3$ spacetime with radius $\ell$.
If we set $M=0$  the metric (\ref{ultraspinning}) provides a parametrization of the AdS$_4$ vacuum with an AdS$_3$ boundary geometry. The asymptotic behaviors (\ref{beha}) are referred to the Fefferman-Graham radial coordinate $r$, which is related to $\rho$ by the condition:
\begin{equation}
\frac{d\rho^2}{1+\frac{\rho^2}{\ell^2}}=\ell^2\,\frac{dr^2}{r^2}\,\,\Rightarrow\,\,\,\,\rho=\frac{r}{2}\,\left(1-\frac{\ell^2}{r^2}\right)\,.
\end{equation}
The reader can verify that, for the vacuum solution $M=0$, the following asymptotic expansions hold:
\begin{align}
V^i&=\frac{r}{2\ell}\,E^i\,\left[1+\frac{\ell^2}{r^2}+O\left(\frac{\ell^6}{r^6}\right)\right]\,,\nonumber\\
\omega^{3i}&=-\frac{r}{2\ell}\,E^i\,\left[1-\frac{\ell^2}{r^2}+O\left(\frac{\ell^6}{r^6}\right)\right]\,,
\end{align}
which yield asymptotic behaviors of $K^i_\pm$ in (\ref{beha}). Note that there is no $O\left(\frac{\ell^3}{r^3}\right)$  term in the expansion of $V^i$, consistently with our assumptions in Section~\ref{asymptoticlimit}. This, together with the absence of $O\left(\left(\frac{\ell}{r}\right)^{\frac{3}{2}}\right)$-terms in the expansion of the gravitinos implies the vanishing of the energy momentum tensor and the supercurrent \cite{Amsel:2009rr,Caldarelli:2012cm}:
\begin{equation}
T_{ij}=0\,\,,\,\,\,J_i=0\,,
\end{equation}
in the boundary CFT.

\section{Proof of eq. (\ref{ansEdet})}\label{proof}
We are going to show here how the parametrization of the super-dreibein $E^i$ on the 3D space-time spanned by the $e^i$:
\begin{eqnarray}\label{ansEdetapp}
E^i &=& \left(1 +\varepsilon\,\frac\ell 2\,  \bar{\chi}\chi -\frac{\ell^2}4 (\bar{\chi}\chi)^2\right) \,e^i\equiv M(\bar \chi \chi) \,e^i\,.
\end{eqnarray}
comes out.
Let us start from eq.s (\ref{d+e}). By plugging the ansatz (\ref{ansE}) into eq. (\ref{d+e}), we get:
\begin{eqnarray}
2 (a + 2 b \bar\chi \chi) \bar\chi \mathcal{D}_{(\varepsilon)}\chi e^i - \frac{\varepsilon}{\ell}\,\epsilon^{ijk}M(\bar \chi \chi)\,e_j\, e_k &=&\epsilon^{ijk} e_j \,e_k\left[\frac 12\bar\chi\chi -\frac{\varepsilon}{\ell}(M(\bar \chi \chi))^2\right]\,.\label{d+E'}
\end{eqnarray}
To solve eq. (\ref{d+E'}) requires   to first determine $\bar\chi\mathcal{D}_{(\varepsilon)}\chi$.
This is found by looking at (\ref{d+chie}), that can be written:
\begin{eqnarray}
\gamma_i(\mathcal{D}_{(\varepsilon)}\chi_A)e^i-\frac \varepsilon\ell\epsilon_{ijk} \gamma_k\chi_A e^i e^j-\frac \varepsilon{2\ell}A \epsilon_{AB}\gamma_i\chi_B e^i&=&0\,,
\end{eqnarray}
implying:
\begin{eqnarray} \label{dchiij}
\gamma_{[i}\mathcal{D}_{(\varepsilon)j]}\chi_A+\frac \varepsilon\ell\epsilon_{ijk} \gamma_k\chi_A  -\frac \varepsilon{2\ell}A_{[j} \epsilon_{AB}\gamma_{i]}\chi_B  &=& 0\,,
\end{eqnarray}
 Multiplication of (\ref{dchiij}) by $\gamma^{ij}$ gives the Dirac equation for $\chi_A$:
\begin{eqnarray}
 \gamma^i\mathcal{D}_{(\varepsilon)i}\chi_A  -\varepsilon\,\frac{3}{\ell}\ii\chi_A -\frac \varepsilon{2\ell}A_i \epsilon_{AB}\gamma^i\chi_B  &=& 0  \,.\label{dchi-}
\end{eqnarray}
while its multiplication for $\gamma^i$:
\begin{eqnarray}\label{eq'}
 \mathcal{D}_{(\varepsilon)j}\chi_A -4\ii\frac \varepsilon\ell \gamma_j\chi_A  -\frac \varepsilon{2\ell}A_{j} \epsilon_{AB}\chi_B +\gamma_j\left( \gamma^i\mathcal{D}_{(\varepsilon)i}\chi_A -\frac \varepsilon{2\ell}A_i \epsilon_{AB}\gamma^i\chi_B\right)  &=& 0\,.
\end{eqnarray}
Combining (\ref{dchi-}) with (\ref{eq'}) we finally get:
\begin{equation}\label{eq}
 \mathcal{D}_{(\varepsilon)j}\chi_A -\ii\frac{\varepsilon}{\ell}   \gamma_j\chi_A  -\frac \varepsilon{2\ell}A_{j} \epsilon_{AB}\chi_B
= 0\,,
\end{equation}

so that, if we now multiply on the left eq. (\ref{eq}) with $\bar \chi_A $, we get
\begin{eqnarray}
\bar \chi \mathcal{D}_{(\varepsilon)j}\chi  &=& 0\,.
\end{eqnarray}

Eq. (\ref{d+E'}) has thus turned  into  an algebraic equation \footnote{Due to the Grassman nature of $\chi_A$, it is in fact a polynomial equation  quadratic in $\bar{\chi}\chi$.}:
\begin{eqnarray}
 - \frac{\varepsilon}{\ell}\,M(\bar \chi \chi)&=&\left[\frac 12 \bar{\chi}\chi -\frac{\varepsilon}{\ell}(M(\bar \chi \chi))^2\right]\,.\label{d+E''}\nonumber\\
 \left(1 +a\,  \bar{\chi}\chi +b(\bar{\chi}\chi)^2\right)&=&\left[1 +(2a-\varepsilon\frac \ell 2)\bar\chi\chi +(a^2+2b)Â§(\bar\chi\chi)^2\right]
\end{eqnarray}
 which is solved
for
$a=\varepsilon \frac\ell 2$, $b= -a^2$, so that (\ref{ansEdet}) follows.


\begin{thebibliography}{99}

 \bibitem{Alvarez:2011gd}
  P.~D.~Alvarez, M.~Valenzuela and J.~Zanelli,
  ``Supersymmetry of a different kind,''
  JHEP {\bf 1204} (2012) 058
  doi:10.1007/JHEP04(2012)058
  [arXiv:1109.3944 [hep-th]].

\bibitem{Maldacena:1997re}
  J.~M.~Maldacena,
  ``The Large N limit of superconformal field theories and supergravity,''
  Adv.\ Theor.\ Math.\ Phys.\  {\bf 2} (1998) 231
  [hep-th/9711200].
\\
  S.~S.~Gubser, I.~R.~Klebanov and A.~M.~Polyakov,
  ``Gauge theory correlators from noncritical string theory,''
  Phys.\ Lett.\ B {\bf 428} (1998) 105
  [hep-th/9802109].
\\
  E.~Witten,
  ``Anti-de Sitter space and holography,''
  Adv.\ Theor.\ Math.\ Phys.\  {\bf 2} (1998) 253
  [hep-th/9802150].
\\
  O.~Aharony, S.~S.~Gubser, J.~M.~Maldacena, H.~Ooguri and Y.~Oz,
  ``Large N field theories, string theory and gravity,''
  Phys.\ Rept.\  {\bf 323} (2000) 183
  [hep-th/9905111].

\bibitem{holren}
  V.~Balasubramanian and P.~Kraus,
  ``A Stress tensor for Anti-de Sitter gravity,''
  Commun.\ Math.\ Phys.\  {\bf 208} (1999) 413
  [hep-th/9902121].
  \\
  J.~de Boer, E.~P.~Verlinde and H.~L.~Verlinde,
  ``On the holographic renormalization group,''
  JHEP {\bf 0008} (2000) 003
  [hep-th/9912012].
\\
  E.~P.~Verlinde and H.~L.~Verlinde,
  ``RG flow, gravity and the cosmological constant,''
  JHEP {\bf 0005} (2000) 034
  [hep-th/9912018].
\\
  J.~de Boer,
  ``The Holographic renormalization group,''
  Fortsch.\ Phys.\  {\bf 49} (2001) 339
  [hep-th/0101026].
 \\
  S.~de Haro, S.~N.~Solodukhin and K.~Skenderis,
  ``Holographic reconstruction of space-time and renormalization in the AdS / CFT correspondence,''
  Commun.\ Math.\ Phys.\  {\bf 217} (2001) 595
  [hep-th/0002230].
  \\
  K.~Skenderis,
  ``Lecture notes on holographic renormalization,''
  Class.\ Quant.\ Grav.\  {\bf 19} (2002) 5849
  [hep-th/0209067].
 \\
For more recent results, see also:
  A.~Lawrence and A.~Sever,
  ``Holography and renormalization in Lorentzian signature,''
  JHEP {\bf 0610} (2006) 013
  [hep-th/0606022].
\\
  I.~Heemskerk and J.~Polchinski,
  ``Holographic and Wilsonian Renormalization Groups,''
  JHEP {\bf 1106} (2011) 031
  [arXiv:1010.1264 [hep-th]].

\bibitem{Guevara:2016rbl}
  A.~Guevara, P.~Pais and J.~Zanelli,
  ``Dynamical Contents of Unconventional Supersymmetry,''
  JHEP {\bf 1608}, 085 (2016)
  doi:10.1007/JHEP08(2016)085
  [arXiv:1606.05239 [hep-th]].

  \bibitem{Alvarez:2013tga}
  P.~D.~Alvarez, P.~Pais and J.~Zanelli,
  ``Unconventional supersymmetry and its breaking,''
  Phys.\ Lett.\ B {\bf 735}, 314 (2014)
  doi:10.1016/j.physletb.2014.06.031
  [arXiv:1306.1247 [hep-th]].

\bibitem{graphene}
A.~H.~Castro Neto, F.~Guinea, N.~M.~R.~Peres, K.~S.~Novoselov and A.~K.~Geim,
  ``The electronic properties of graphene,''
  Rev.\ Mod.\ Phys.\  {\bf 81} (2009) 109.
  doi:10.1103/RevModPhys.81.109.\\
  For a recent review, see for example:\\
  A.~Lucas and K.~C.~Fong,
  ``Hydrodynamics of electrons in graphene,''
  J.\ Phys.\ Condens.\ Matter {\bf 30} (2018) 053001
  doi:10.1088/1361-648X/aaa274
  [arXiv:1710.08425 [cond-mat.str-el]].



  \bibitem{Cortijo:2006ej}
  A.~Cortijo and M.~A.~H.~Vozmediano,
  ``A Cosmological model for corrugated graphene sheets,''
  Eur.\ Phys.\ J.\ ST {\bf 148}, 83 (2007)
  doi:10.1140/epjst/e2007-00228-2
  [cond-mat/0612623].
  \\
  A.~Cortijo and M.~A.~H.~Vozmediano,
  ``Effects of topological defects and local curvature on the electronic properties of planar graphene,''
  Nucl.\ Phys.\ B {\bf 763}, 293 (2007)
  [Nucl.\ Phys.\ B {\bf 807}, 659 (2009)]
  doi:10.1016/j.nuclphysb.2008.09.006, 10.1016/j.nuclphysb.2006.10.031
  [cond-mat/0612374].
  \\
  A.~Cortijo, F.~Guinea and M.~A.~H.~Vozmediano,
  ``Geometrical and topological aspects of graphene and related materials,''
  J.\ Phys.\ A {\bf 45}, 383001 (2012)
  doi:10.1088/1751-8113/45/38/383001
  [arXiv:1112.2054 [cond-mat.mes-hall]].

  \bibitem{Iorio:2011yz}
  A.~Iorio and G.~Lambiase,
  ``The Hawking-Unruh phenomenon on graphene,''
  Phys.\ Lett.\ B {\bf 716}, 334 (2012)
  doi:10.1016/j.physletb.2012.08.023
  [arXiv:1108.2340 [cond-mat.mtrl-sci]].
  \\
  A.~Iorio and G.~Lambiase,
  ``Quantum field theory in curved graphene spacetimes, Lobachevsky geometry, Weyl symmetry, Hawking effect, and all that,''
  Phys.\ Rev.\ D {\bf 90}, no. 2, 025006 (2014)
  doi:10.1103/PhysRevD.90.025006
  [arXiv:1308.0265 [hep-th]].

  \bibitem{Cvetic:2012vg}
  M.~Cvetic and G.~W.~Gibbons,
  ``Graphene and the Zermelo Optical Metric of the BTZ Black Hole,''
  Annals Phys.\  {\bf 327}, 2617 (2012)
  doi:10.1016/j.aop.2012.05.013
  [arXiv:1202.2938 [hep-th]].

  \bibitem{Andrianopoli:2014aqa}
  L.~Andrianopoli and R.~D'Auria,
  ``N=1 and N=2 pure supergravities on a manifold with boundary,''
  JHEP {\bf 1408} (2014) 012
  doi:10.1007/JHEP08(2014)012
  [arXiv:1405.2010 [hep-th]].

\bibitem{Amsel:2009rr}
  A.~J.~Amsel and G.~Comp\`ere,
``Supergravity at the boundary of AdS supergravity,''
  Phys.\ Rev.\ D {\bf 79} (2009) 085006
  doi:10.1103/PhysRevD.79.085006
  [arXiv:0901.3609 [hep-th]].

 \bibitem{Caldarelli:2008pz}
  M.~M.~Caldarelli, R.~Emparan and M.~J.~Rodriguez,
 ``Black Rings in (Anti)-deSitter space,''
  JHEP {\bf 0811} (2008) 011
  doi:10.1088/1126-6708/2008/11/011
  [arXiv:0806.1954 [hep-th]].

\bibitem{Caldarelli:2012cm}
  M.~M.~Caldarelli, R.~G.~Leigh, A.~C.~Petkou, P.~M.~Petropoulos, V.~Pozzoli and K.~Siampos,
  ``Vorticity in holographic fluids,''
  PoS CORFU {\bf 2011} (2011) 076
  [arXiv:1206.4351 [hep-th]].

\bibitem{Gnecchi:2013mja}
  A.~Gnecchi, K.~Hristov, D.~Klemm, C.~Toldo and O.~Vaughan,
  ``Rotating black holes in 4d gauged supergravity,''
  JHEP {\bf 1401}, 127 (2014)
  doi:10.1007/JHEP01(2014)127
  [arXiv:1311.1795 [hep-th]].

  \bibitem{Klemm:2014rda}
  D.~Klemm,
  ``Four-dimensional black holes with unusual horizons,''
  Phys.\ Rev.\ D {\bf 89} (2014) no.8,  084007
  doi:10.1103/PhysRevD.89.084007
  [arXiv:1401.3107 [hep-th]].

  \bibitem{Hennigar:2015cja}
  R.~A.~Hennigar, D.~Kubiz\v{n}\'ak, R.~B.~Mann and N.~Musoke,
  ``Ultraspinning limits and super-entropic black holes,''
  JHEP {\bf 1506} (2015) 096
  doi:10.1007/JHEP06(2015)096
  [arXiv:1504.07529 [hep-th]].

 \bibitem{Achucarro:1987vz}
  A.~Achucarro and P.~K.~Townsend,
  ``A Chern-Simons Action for Three-Dimensional anti-De Sitter Supergravity Theories,''
  Phys.\ Lett.\ B {\bf 180} (1986) 89.
  doi:10.1016/0370-2693(86)90140-1

  \bibitem{Achucarro:1989gm}
  A.~Achucarro and P.~K.~Townsend,
  ``Extended Supergravities in $d$ = (2+1) as {Chern-Simons} Theories,''
  Phys.\ Lett.\ B {\bf 229} (1989) 383.
  doi:10.1016/0370-2693(89)90423-1

  \bibitem{Witten:1988hc}
  E.~Witten,
  ``(2+1)-Dimensional Gravity as an Exactly Soluble System,''
  Nucl.\ Phys.\ B {\bf 311}, 46 (1988).
  doi:10.1016/0550-3213(88)90143-5

  \bibitem{Witten:2007kt}
  E.~Witten,
  ``Three-Dimensional Gravity Revisited,''
  arXiv:0706.3359 [hep-th].

  \bibitem{Witten:2003ya}
  E.~Witten,
  ``SL(2,Z) action on three-dimensional conformal field theories with Abelian symmetry,''
  In *Shifman, M. (ed.) et al.: From fields to strings, vol. 2* 1173-1200
  [hep-th/0307041].

\bibitem{Cacciatori:2005wz}
  S.~L.~Cacciatori, M.~M.~Caldarelli, A.~Giacomini, D.~Klemm and D.~S.~Mansi,
  ``Chern-Simons formulation of three-dimensional gravity with torsion and nonmetricity,''
  J.\ Geom.\ Phys.\  {\bf 56}, 2523 (2006)
  doi:10.1016/j.geomphys.2006.01.006
  [hep-th/0507200].

 \bibitem{Leigh:2008tt}
  R.~G.~Leigh, N.~N.~Hoang and A.~C.~Petkou,
  ``Torsion and the Gravity Dual of Parity Symmetry Breaking in AdS(4) / CFT(3) Holography,''
  JHEP {\bf 0903}, 033 (2009)
  doi:10.1088/1126-6708/2009/03/033
  [arXiv:0809.5258 [hep-th]].

\bibitem{Townsend:2013ela}
  P.~K.~Townsend and B.~Zhang,
  ``Thermodynamics of ``Exotic'' Ba\~nados-Teitelboim-Zanelli Black Holes,''
  Phys.\ Rev.\ Lett.\  {\bf 110}, no. 24, 241302 (2013)
  doi:10.1103/PhysRevLett.110.241302
  [arXiv:1302.3874 [hep-th]].

\bibitem{York:1972sj}
  J.~W.~York, Jr.,
  ``Role of conformal three geometry in the dynamics of gravitation,''
  Phys.\ Rev.\ Lett.\  {\bf 28} (1972) 1082.
 \\
  J.~D.~Brown and J.~W.~York, Jr.,
  ``Quasilocal energy and conserved charges derived from the gravitational action,''
  Phys.\ Rev.\ D {\bf 47} (1993) 1407
  [gr-qc/9209012].

\bibitem{Gibbons:1976ue}
  G.~W.~Gibbons and S.~W.~Hawking,
  ``Action Integrals and Partition Functions in Quantum Gravity,''
  Phys.\ Rev.\ D {\bf 15} (1977) 2752.

\bibitem{Horava:1996ma}
  P.~Horava and E.~Witten,
  ``Eleven-dimensional supergravity on a manifold with boundary,''
  Nucl.\ Phys.\ B {\bf 475} (1996) 94
  [hep-th/9603142].

\bibitem{Aros:1999id}
  R.~Aros, M.~Contreras, R.~Olea, R.~Troncoso and J.~Zanelli,
  ``Conserved charges for gravity with locally AdS asymptotics,''
  Phys.\ Rev.\ Lett.\  {\bf 84} (2000) 1647
  [gr-qc/9909015].

 \bibitem{Miskovic:2009bm}
  O.~Miskovic and R.~Olea,
  ``Topological regularization and self-duality in four-dimensional anti-de Sitter gravity,''
  Phys.\ Rev.\ D {\bf 79} (2009) 124020
  doi:10.1103/PhysRevD.79.124020
  [arXiv:0902.2082 [hep-th]].

\bibitem{vanNieuwenhuizen:2005kg}
  P.~van Nieuwenhuizen and D.~V.~Vassilevich,
  ``Consistent boundary conditions for supergravity,''
  Class.\ Quant.\ Grav.\  {\bf 22} (2005) 5029
  [hep-th/0507172].
\\
  D.~V.~Belyaev,
  ``Boundary conditions in supergravity on a manifold with boundary,''
  JHEP {\bf 0601} (2006) 047
  [hep-th/0509172].
\\
  D.~V.~Belyaev and P.~van Nieuwenhuizen,
  ``Tensor calculus for supergravity on a manifold with boundary,''
  JHEP {\bf 0802} (2008) 047
  [arXiv:0711.2272 [hep-th]].
\\
  D.~V.~Belyaev and P.~van Nieuwenhuizen,
  ``Simple d=4 supergravity with a boundary,''
  JHEP {\bf 0809} (2008) 069
  [arXiv:0806.4723 [hep-th]].
\\
  D.~Grumiller and P.~van Nieuwenhuizen,
  ``Holographic counterterms from local supersymmetry without boundary conditions,''
  Phys.\ Lett.\ B {\bf 682} (2010) 462
  [arXiv:0908.3486 [hep-th]].
  D.~V.~Belyaev and T.~G.~Pugh,
  ``The Supermultiplet of boundary conditions in supergravity,''
  JHEP {\bf 1010} (2010) 031
  [arXiv:1008.1574 [hep-th]].

\bibitem{esposito}
  G.~Esposito, A.~Y.~.Kamenshchik and K.~Kirsten,
  ``One loop effective action for Euclidean Maxwell theory on manifolds with boundary,''
  Phys.\ Rev.\ D {\bf 54} (1996) 7328
  [hep-th/9606132].
  \\
  I.~G.~Avramidi and G.~Esposito,
  ``Gauge theories on manifolds with boundary,''
  Commun.\ Math.\ Phys.\  {\bf 200} (1999) 495
  [hep-th/9710048].

\bibitem{Moss:2003bk}
  I.~G.~Moss,
  ``Boundary terms for eleven-dimensional supergravity and M theory,''
  Phys.\ Lett.\ B {\bf 577} (2003) 71
  [hep-th/0308159].
\\
  I.~GMoss,
  ``Boundary terms for supergravity and heterotic M theory,''
  Nucl.\ Phys.\ B {\bf 729} (2005) 179
  [hep-th/0403106].

\bibitem{Howe:2011tm}
  P.~S.~Howe, T.~G.~Pugh, K.~S.~Stelle and C.~Strickland-Constable,
  ``Ectoplasm with an Edge,''
  JHEP {\bf 1108} (2011) 081
  [arXiv:1104.4387 [hep-th]].

\bibitem{Freedman:2016yue}
  D.~Z.~Freedman, K.~Pilch, S.~S.~Pufu and N.~P.~Warner,
  ``Boundary Terms and Three-Point Functions: An AdS/CFT Puzzle Resolved,''
  JHEP {\bf 1706} (2017) 053
  doi:10.1007/JHEP06(2017)053
  [arXiv:1611.01888 [hep-th]].

 \bibitem{Genolini:2016ecx}
  P.~Benetti Genolini, D.~Cassani, D.~Martelli and J.~Sparks,
  ``Holographic renormalization and supersymmetry,''
  JHEP {\bf 1702} (2017) 132
  doi:10.1007/JHEP02(2017)132
  [arXiv:1612.06761 [hep-th]].

\bibitem{Drukker:2017xrb}
  N.~Drukker, D.~Martelli and I.~Shamir,
  ``The energy-momentum multiplet of supersymmetric defect field theories,''
  JHEP {\bf 1708} (2017) 010
  doi:10.1007/JHEP08(2017)010
  [arXiv:1701.04323 [hep-th]].

\bibitem{Aros:1999kt}
  R.~Aros, M.~Contreras, R.~Olea, R.~Troncoso and J.~Zanelli,
  ``Conserved charges for even dimensional asymptotically AdS gravity theories,''
  Phys.\ Rev.\ D {\bf 62} (2000) 044002
  [hep-th/9912045].

\bibitem{Mora:2004kb}
  P.~Mora, R.~Olea, R.~Troncoso and J.~Zanelli,
  ``Finite action principle for Chern-Simons AdS gravity,''
  JHEP {\bf 0406} (2004) 036
  [hep-th/0405267].

\bibitem{Olea:2005gb}
  R.~Olea,
  ``Mass, angular momentum and thermodynamics in four-dimensional Kerr-AdS black holes,''
  JHEP {\bf 0506} (2005) 023
  [hep-th/0504233].

\bibitem{Jatkar:2014npa}
  D.~P.~Jatkar, G.~Kofinas, O.~Miskovic and R.~Olea,
  ``Conformal Mass in AdS gravity,''
  arXiv:1404.1411 [hep-th].

   \bibitem{Banados:1992wn}
  M.~Ba\~nados, C.~Teitelboim and J.~Zanelli,
  ``The Black hole in three-dimensional space-time,''
  Phys.\ Rev.\ Lett.\  {\bf 69}, 1849 (1992)
  doi:10.1103/PhysRevLett.69.1849
  [hep-th/9204099].

  \bibitem{Banados:1992gq}
  M.~Ba\~nados, M.~Henneaux, C.~Teitelboim and J.~Zanelli,
  ``Geometry of the (2+1) black hole,''
  Phys.\ Rev.\ D {\bf 48}, 1506 (1993)
  Erratum: [Phys.\ Rev.\ D {\bf 88}, 069902 (2013)]
  doi:10.1103/PhysRevD.48.1506, 10.1103/PhysRevD.88.069902
  [gr-qc/9302012].

\bibitem{Emparan:1999fd}
  R.~Emparan, G.~T.~Horowitz and R.~C.~Myers,
``Exact description of black holes on branes. 2. Comparison with BTZ black holes and black strings,''
  JHEP {\bf 0001} (2000) 021
  doi:10.1088/1126-6708/2000/01/021
  [hep-th/9912135].



  \bibitem{Semenoff:2012xu}
  G.~W.~Semenoff,
  ``Engineering holographic graphene,''
  AIP Conf.\ Proc.\  {\bf 1483} (2012) 305.
  doi:10.1063/1.4756976


  \end{thebibliography}
\end{document}